\documentclass[conference]{IEEEtran}
\IEEEoverridecommandlockouts

\usepackage{amsmath,graphicx}
\usepackage{multirow, booktabs}
\usepackage{pifont}
\usepackage{unicode}
\usepackage{xcolor}
\usepackage[numbers,sort&compress]{natbib}
\usepackage{enumitem}

\newcommand{\texttoken}[1]{{\small\textsf{#1}}}
\newcommand{\beep}{\textbf{\textit{beep}}}

\title{\fontsize{22}{26}\selectfont META-CAT: Speaker-Informed Speech Embeddings via Meta Information Concatenation for Multi-talker ASR}
\author{
\begin{tabular}{c} 
Jinhan Wang, Weiqing Wang, Kunal Dhawan, Taejin Park, Myungjong Kim, Ivan Medennikov, \\ 
He Huang, Nithin Koluguri, Jagadeesh Balam and Boris Ginsburg\\
\textit{NVIDIA, Santa Clara, CA, USA}\\
\end{tabular} 
}

\begin{document}

\maketitle

\begin{abstract}
We propose a novel end-to-end multi-talker automatic speech recognition (ASR) framework that enables both multi-speaker (MS) ASR and target-speaker (TS) ASR. Our proposed model is trained in a fully end-to-end manner, incorporating speaker supervision from a pre-trained speaker diarization module. We introduce an intuitive yet effective method for masking ASR encoder activations using output from the speaker supervision module, a technique we term Meta-Cat (meta-information concatenation), that can be applied to both MS-ASR and TS-ASR. Our results demonstrate that the proposed architecture achieves competitive performance in both MS-ASR and TS-ASR tasks, without the need for traditional methods, such as neural mask estimation or masking at the audio or feature level. Furthermore, we demonstrate a glimpse of a unified dual-task model which can efficiently handle both MS-ASR and TS-ASR tasks. Thus, this work illustrates that a robust end-to-end multi-talker ASR framework can be implemented with a streamlined architecture, obviating the need for the complex speaker filtering mechanisms employed in previous studies.
\end{abstract}

\section{Introduction}
\label{sec:intro}

Recent advancements in ASR, driven by improved architectures and larger training datasets, have significantly advanced the field. Concurrently, interest in multi-talker ASR has grown, particularly for applications such as analyzing natural conversations, developing voice assistants, and transcribing speech in health and legal contexts. Although this task is referred to by various terms, such as speaker-attributed ASR, we define two primary multi-talker ASR systems: multi-speaker (MS) ASR and target-speaker (TS) ASR. Both tasks require attention to speaker-specific information during transcription, but the key distinction is that MS-ASR aims to transcribe speech from all speakers in an audio mixture, whereas TS-ASR focuses exclusively on transcribing the utterances of a specific speaker using a short reference sample of their voice.

Typically, these two tasks are approached using cascade systems. In MS-ASR, a speaker diarization system is often used to separate the speech of each individual speaker, after which a single-speaker ASR model transcribes the corresponding segments \cite{medennikov2020stc}, using guided source separation (GSS) to estimate spectral masks~\cite{boeddeker2018front}. Alternatively, speaker-attributed ASR systems employ speaker embeddings to simultaneously generate transcriptions for all speakers in the mixture~\cite{kanda2022transcribe}. Recently, Cornell et al.~\cite{cornell2024one} introduced a joint system that integrates speaker diarization and ASR, performing speaker clustering as part of the MS-ASR process. Similarly, TS-ASR is viewed as a combination of a speech enhancement front-end, in~\cite{moriya2022streaming}. 

\begin{figure}
\centering
\includegraphics[width=0.49\textwidth]{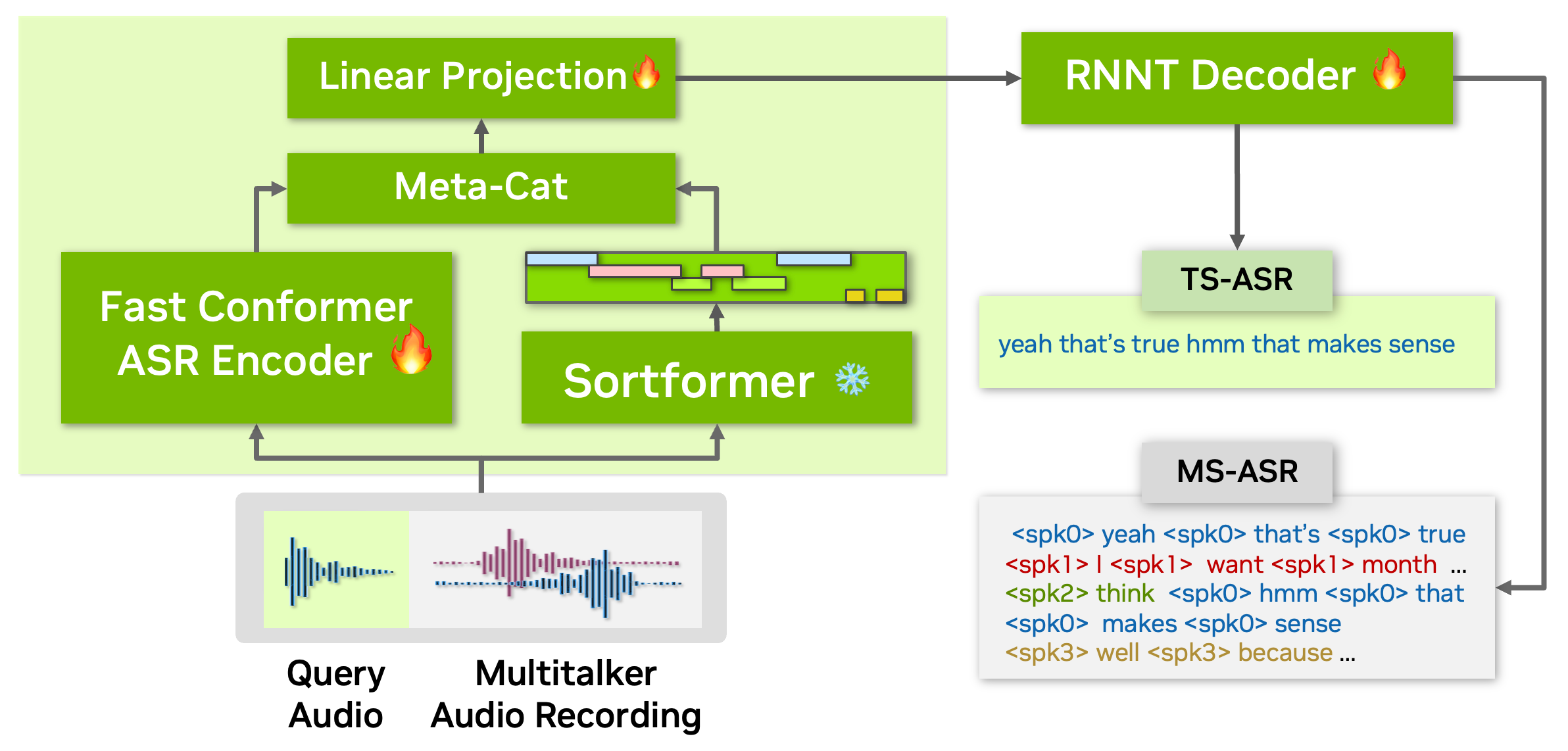}
\vspace{-4ex}
\caption{Data flow of the proposed system that supports both MS-ASR and TS-ASR.}
\label{fig:main_dataflow}
\vspace{-4ex}
\end{figure}

In end-to-end systems, the design of input/output representations and training objectives becomes critical. For MS-ASR, permutation invariant training (PIT) was introduced to address the challenge of varying speaker order during training \cite{yu17b_INTERSPEECH}, though it can struggle in scenarios requiring speaker tracking \cite{kanda19_interspeech}. A significant advancement came with Kanda et al.'s~\cite{kanda2020serialized} Serialized Output Training (SOT), which simplified the training of end-to-end MS-ASR systems. The SOT method processes overlapping speech using multi-head self-attention mechanisms and was later extended to a token-level version (t-SOT) for streaming applications \cite{kanda2022streaming}. For end-to-end TS-ASR, an additional speaker encoder is typically required to extract a speaker profile from a short sample of the target speaker’s voice, allowing the system to mask or filter out the speech of other speakers. The speaker encoder is often a frozen, pre-trained speaker recognition model \cite{moriya2022streaming, kanda19_interspeech, zhang2023conformer}. Most recently, a single model that conditionally performs either MS-ASR or TS-ASR was proposed in~\cite{masumura2023end}.

In this paper, we propose an end-to-end framework that performs both MS-ASR and TS-ASR using a unified architecture. The major contributions of this work are:
\begin{itemize}[left=0.2cm] 
    \item The proposed framework utilizes speaker supervision from an end-to-end speaker diarization model Sortformer~\cite{park2024sortformer}, without employing spectral mask estimation~\cite{boeddeker2018front, medennikov2020stc} or speaker embeddings~\cite{vzmolikova2019speakerbeam,zhang2023conformer}.
    \item We propose a novel speaker information encoding scheme, Meta-Cat, for injecting speaker supervision information into ASR embeddings in a generic manner to tackle both MS-ASR and TS-ASR tasks with an identical architecture. 
    \item We propose task-specific end-to-end models for MS-ASR and TS-ASR based on identical architecture. In addition, we demonstrate a dual-task model that can perform MS-ASR and TS-ASR simultaneously depending on the query input. 
\end{itemize}

\section{Proposed System}

\subsection{Sortformer Diarizer in End-to-End Multi-talker ASR System}
We employ \textit{Sortformer}~\cite{park2018multimodal}, an end-to-end speaker diarization model that generates speaker timestamps in arrival-time order (ATO). On the ASR side, we employ a pre-trained FastConformer~\cite{rekesh2023fast} encoder module that processes the acoustic signal and provides ASR embeddings (ASR encoder states). Subsequently, the Meta-Cat module injects the estimated speaker-annotated timestamps into ASR embeddings. We refer to the speaker-annotated timestamps as \textit{speaker supervision}. The processed ASR encoder state sequences embedded with speaker supervision information are then fed into a recurrent neural network transducer (RNNT) decoder~\cite{graves2012sequence}. Finally, the RNNT decoder generates speaker tokens with transcription, as described in Fig.~\ref{fig:main_dataflow}. 

For MS-ASR, the resulting transcription follows the format shown in Fig.~\ref{fig:main_dataflow}, where each speaker's transcription is preceded by a speaker token. Following the speech segments in ATO, the speaker token indices are also assigned according to the ATO as shown in Fig.~\ref{fig:ms_and_ts_asr}. We also leverage this ability to order speech segments by arrival time to perform the TS-ASR task.

\subsection{Multi-Speaker ASR} 
\label{sec:msasr}
Fig.~\ref{fig:ms_and_ts_asr}-(a) illustrates an example of our proposed end-to-end MS-ASR. For MS-ASR models, we follow the configurations in \cite{park2024sortformer}, regarding the arrangement of word-level speaker tokens for training. For the dual-task model, if query audio is not provided and there is no trigger signal for TS-ASR, the model automatically performs the MS-ASR task.

\subsection{Target-Speaker ASR}
\begin{figure}[!tbp]
\centering
\includegraphics[width=0.48\textwidth]{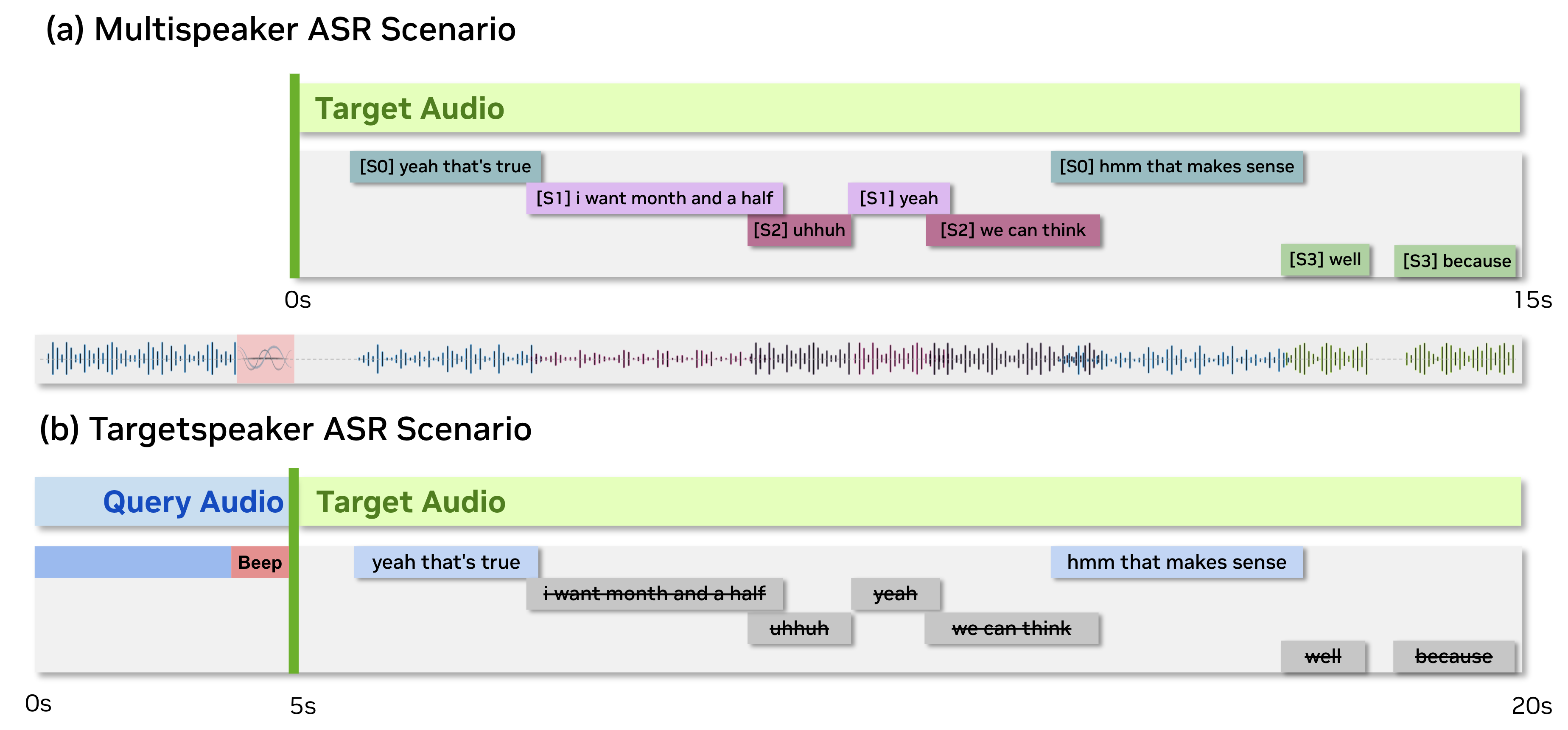}
\vspace{-2ex}
\caption{Exemplary illustration of MS-ASR and TS-ASR.}
\label{fig:ms_and_ts_asr}
\vspace{-2ex}
\end{figure}

As described in \cite{park2024sortformer}, Sortformer assigns the speaker token \texttoken{<|spk0|>} to the first occurring speaker's utterances. Thus, TS-ASR can be easily extended from MS-ASR by prepending a query audio segment before the target audio to mark the query speaker
as \texttoken{<|spk0|>}. Most importantly, we utilize a trigger signal for the TS-ASR task to inform the model the endpoint of the query audio segment and target audio is starting. As illustrated in Fig.\ref{fig:ms_and_ts_asr}-(b), we refer to the trigger signal as \beep. Accordingly, only target speaker's groundtruth transcription in target audio is provided for training, and we expect the target-speaker's speech to be transcribed during inference.

\subsection{Speaker Supervision Sources}
Speaker supervision is a stream of speaker label information from either groundtruth, which is sourced from a Rich Transcription Time Marked (RTTM)~\cite{sun2010speaker} file, or diarization predictions from Sortformer model. We denote this speaker supervision type as \textit{RTTM}. On the contrary, the other speaker supervision strategy that uses pure diarization output is denoted as \textit{DIAR}. 

\subsection{Speaker Encoding Schemes}
\subsubsection{Meta-Cat}
\begin{figure}[tbp!]
\centering
\includegraphics[width=0.45\textwidth]{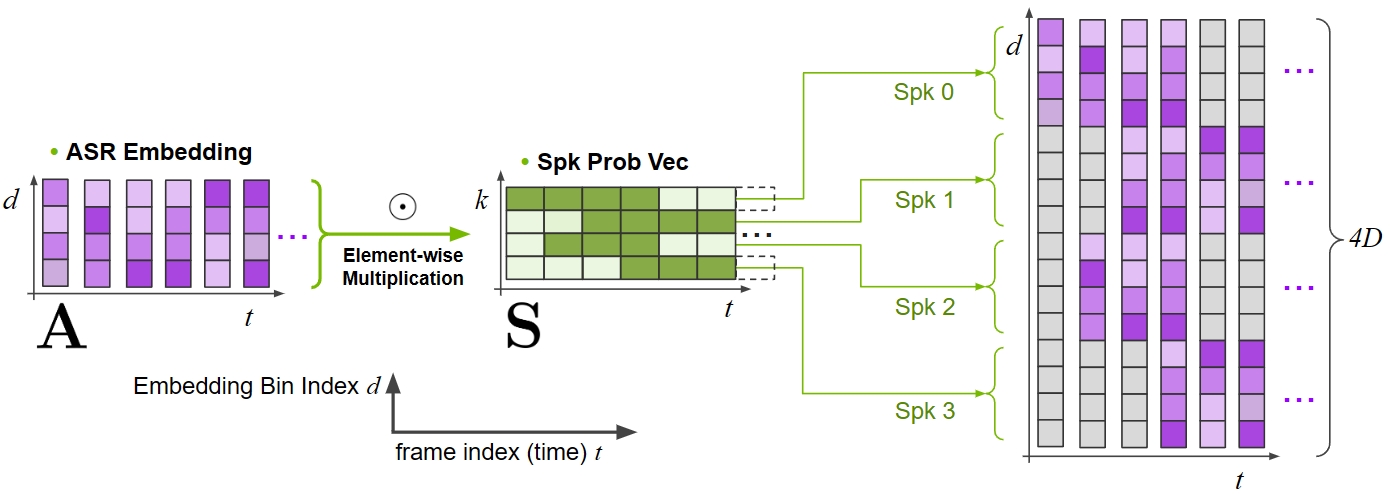}
\vspace{-1ex}
\caption{Meta-Cat converts the predicted timestamps to a masked embedding sequence that conveys speaker supervision information.}
\vspace{-2ex}
\label{fig:metacat}
\end{figure}

The core ingredient of our proposed end-to-end model is the speaker encoding, which controls how speaker information is injected into the ASR embeddings. Our proposed method, Meta-Cat, expands the pure ASR embedding into multiple sub-spaces, triggered by the speaker probability vector, which are then concatenated to be a super-vector. As shown in Fig.~\ref{fig:metacat}, we denote the ASR embedding as $\textbf{A} \in \mathbb{R}^{D \times T}$, and speaker probability vector as  $\textbf{S} \in \mathbb{R}^{K \times T}$, where \textit{D} is the hidden dimension of ASR embeddings, $K$ is the number of speakers, and $T$ is the embedding hidden length for both \textbf{A} and \textbf{S}. 

\begin{figure}[tbp!]
\centering
\includegraphics[width=0.48\textwidth]{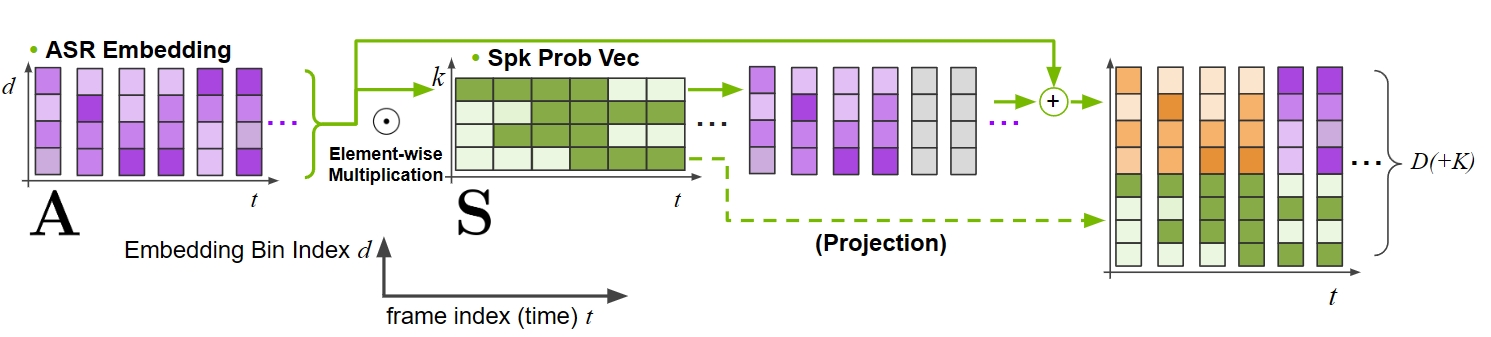}

\caption{Meta-Cat-Residual (+Projection): Adding residual connections and another projection layer on top of Meta-Cat.}
\label{fig:MCR}
\vspace{-3ex}
\end{figure}

\subsubsection{Meta-Cat-Residual}
TS-ASR requires only one subspace $\textit{C}_0$ being activated for the target speaker. However, Meta-Cat expands the ASR embedding multiple times according to the number of speakers and it might introduce redundancy. 
\textit{Meta-Cat-Residual (Meta-Cat-R)} is proposed to alleviate the redundancy as shown in Fig.~\ref{fig:MCR}. During speaker encoding, only the first row corresponding to \texttoken{<|spk0|>} is extracted and multiplied by the ASR embedding to obtain $\textit{C}_0$. Furthermore, to deal with information loss due to inaccurate speaker probability vector when using DIAR, the resulting \texttoken{<|spk0|>}'s ASR embedding  $\textit{C}_0$ is then skip-connected with the original ASR embedding \textbf{A} in a residual manner.

\subsubsection{Meta-Cat-Residual-Projection}
Preliminary experiments showed that the model can handle the inaccurate speaker probability vectors by recovering the correct speaker mapping, for instance, \texttoken{<|spk1|>} $\rightarrow$ \texttoken{<|spk0|>}. Therefore, distinguishable patterns across non-target speakers might still be necessary for the TS-ASR task. Thus, \textit{Meta-Cat-Residual-Projection (Meta-Cat-RP)} concatenates the speaker probability vector with the embedding obtained through Meta-Cat-R as shown in Fig.~\ref{fig:MCR}. With the additional concatenated speaker probability vector, the model preserves the knowledge of all speakers' diarization information but still makes target-speaker stand out. 

\subsection{Unified Dual-Task Model: MS-ASR + TS-ASR}
A dual-task model that can perform dual tasks, MS-ASR and TS-ASR simultaneously can be accomplished. First, the trigger signal (\beep) and corresponding token (\texttoken{<beep>}) are injected regardless of whether a query audio is provided, as in a TS-ASR task. Second, the dual-task model is trained in an MS-ASR manner, where the groundtruth transcriptions from all speakers in target audio are provided. When a query audio is provided, the target speaker's transcription can be obtained from the transcriptions associated with \texttoken{<|spk0|>} in the dual-task model while performing MS-ASR.

\section{Experimental Results}
\label{sec:exp}

\subsection{Model Training} 

\subsubsection{Dataset} 
\label{sec:dataset}

For training, we use AMI \cite{ami} Individual Headset Mix (IHM) train split appeared in~\cite{landini2022bayesian}, all sessions in  ICSI~\cite{icsi}, and DipCo~\cite{dipco} dataset. The three sets collectively contain 138 hours of multi-speaker speech, with up to four speakers per sample. 
For TS-ASR model training, the duration of the query audio ranges from 3 to 10 seconds. To ensure the query audio contains sufficient target-speaker acoustic information, the query audio is constrained to have at least 5 words. For LibriSpeechMix~\cite{kanda2020serialized}, we create a 960-hour training set from LibriSpeech~\cite{panayotov2015librispeech}, following the recipe in LibriSpeechMix~\cite{kanda2020serialized}.

\subsubsection{Data Cleaning} For the data-cleaning processs of MS-ASR, we follow the identical process in \cite{park2024sortformer}. For TS-ASR pipeline, transcription samples are generated by extracting all sentences following \texttoken{<|spk0|>} and merging them sequentially. Note that there will be no query audio transcription or speaker tokens included in the groundtruth transcription.

For the unified MS+TS-ASR system, the general procedure is similar to MS-ASR, but the first speaker will be determined by the optional query speaker. The first speaker token will be \texttoken{<|spk0|>} if this speaker is the same as the query speaker or if no query audio is provided; otherwise, the transcription starts with \texttoken{<|spk1|>}.

\subsubsection{Pre-trained Model} 
We employ a pre-trained sortformer diarization model proposed in \cite{park2024sortformer}. The sortformer diarization model is only used when speaker supervision scheme is specified as \textit{DIAR}, and the sortformer models are frozen during training. For the pre-trained ASR model, we use Parakeet-RNNT model~\cite{nvidia_parakeet_rnnt} based on NeMo framework~\cite{nemo}. We finetune both ASR encoder and decoder.

\subsubsection{Evaluation} 
\label{sec:evaluation}
For MS-ASR, word error rate (WER) and concatenated minimum-permutation word error rate (cpWER)~\cite{watanabe2020chime} are used to evaluate our proposed models. 
The cpWER is calculated by combining each speaker's utterances separately and finding the lowest WER for different speaker combinations. It reflects errors from both speech recognition and speaker diarization. 
For TS-ASR, the WER is used to evaluate how the model performs in transcribing only the target-speaker's transcription regardless of the other speakers for the AMI dataset. We also used the TS-WER mentioned in \cite{zhang2023conformer} for LibriSpeechMix evaluation. 

For dual-task model evaluation, MS-ASR task is evaluated using WER and cpWER across all samples. TS-ASR task is evaluated across samples with query audio prepended using WER by extracting the target-speaker transcription from both groundtruth and prediction. 

For the MS-ASR, the model is trained on the training dataset and tested on both AMI-IHM test split appeared in~\cite{landini2022bayesian} and CH109, which is a two-speaker subset of 109 session from Callhome American English Speech~\cite{canavan1997CALLHOME}. The TS-ASR models are trained on training set and tested on both AMI-test and LibrSpeechMix~\cite{kanda2020serialized}.

\label{sec:results}
\subsection{Multi-Speaker ASR}
\subsubsection{Speaker Encoding Schemes}
\newcommand{\cmark}{\ding{51}}%
\newcommand{\xmark}{\ding{55}}%
\begin{table}[tbp!]
\caption{MS-ASR model evaluations on 10s-20s AMI-IHM test set }
  \label{tab:spk_kernel_multi}
  \centering
  \vspace{-2ex}
  \begin{tabular}[c]{@{\hspace{0.85ex}}r@{\hspace{0.85ex}}c@{\hspace{0.85ex}}c@{\hspace{0.85ex}}|@{\hspace{0.85ex}}c@{\hspace{0.85ex}}c@{\hspace{0.85ex}}|@{\hspace{0.85ex}}c@{\hspace{0.85ex}}c@{\hspace{0.85ex}}|@{\hspace{0.85ex}}c@{\hspace{0.85ex}}c@{\hspace{0.85ex}}}
    \toprule
     \textbf{Speaker} & \multicolumn{4}{c}{\textbf{AMI}} & \multicolumn{4}{c}{\textbf{CH109}}
     \\ 
     \textbf{Encoding}& \multicolumn{2}{c}{RTTM} & \multicolumn{2}{c}{DIAR} & \multicolumn{2}{c}{RTTM} & \multicolumn{2}{c}{DIAR}  \\ 
     \textbf{Scheme}& WER & cpWER & WER & cpWER & WER & cpWER  & WER & cpWER \\
     
   \midrule
        No Spk-Sup.  & 21.29 & 29.04  & - & -  & 20.32 & 30.26  & - & - \\
	Projection & 21.05 & 25.06 & 21.33 & 26.83  & 20.52 & 29.09 & 26.34 & 27.50 \\
	Sinusoidal & 20.80 & 25.64  & \textbf{20.87} & 26.75  & \textbf{20.20} & 28.55 & 25.87 & 28.92 \\
	Meta-Cat & \textbf{20.15} & \textbf{22.83} & \textbf{20.87} & \textbf{26.02} & 25.48 & \textbf{28.41} & \textbf{25.74} & \textbf{26.17}  \\
 \bottomrule
   \vspace{-5ex}
     \end{tabular}
\end{table}

     

Table~\ref{tab:spk_kernel_multi} presents the results of different speaker encoding schemes applied to MS-ASR tasks on the AMI and CH109 datasets, evaluated using both RTTM and DIAR supervision. The Meta-Cat method outperforms the baseline (No Spk-Sup.: No Speaker Supervision) and other encoding methods, achieving the lowest WER and cpWER across both datasets in most scenarios. On the AMI test set, Meta-Cat achieves a relative WER reduction of 21.4\% under RTTM and 10.5\% under DIAR, demonstrating its robustness in incorporating speaker information. Sinusoidal encoding also performs well, especially with DIAR on AMI, showing competitive results close to Meta-Cat.

\subsubsection{RTTM-Mix-Prob}
To investigate the encoding robustness toward inaccurate speaker probability vectors from the diarization model, we compared the performance using different ratios of RTTM and DIAR speaker supervision during training, where the ratio is defined as RTTM-Mix-Prob. If the ratio is set to 0.5, then at each time step for each sample, the speaker probability vector has a 50\% chance of having RTTM instead of diarization output for all speakers' components.

\begin{table}[tbp!]
\caption{MS-ASR model with Meta-Cat over various RTTM-Mix-Prob }
\vspace{-3ex}
  \label{tab:metacat_mix_prob}
  \centering
  \vspace{2ex}
  \begin{tabular}[c]{ccc|cc}
    \toprule
     \multirow{2}*{\textbf{RTTM-Mix-Prob}}  & \multicolumn{2}{c}{RTTM} & \multicolumn{2}{c}{DIAR}  \\ 
     & WER & cpWER & WER & cpWER\\
     
   \midrule
        0 (DIAR)  & 20.78 & 28.10 & 20.76 & 26.86\\
	0.25 & 20.79 & 26.73 & 20.80 & 26.86 \\
	0.5 & 20.78 & 25.78 & 20.91 & 26.62 \\
	0.75 & 20.43 & 24.75 & \textbf{20.60} & 26.40\\
    1 (RTTM) & \textbf{20.15} & \textbf{22.83} & 20.87 & \textbf{26.02}\\
        \bottomrule
     \end{tabular}
     \vspace{-2ex}
\end{table}

As shown in Table~\ref{tab:metacat_mix_prob}, as more RTTM speaker supervision is included in model training, better cpWER is obtained. Larger improvements are observed using RTTM during evaluation (\textasciitilde 1\% absolute per step from 0 to 1 in RTTM-Mix-Prob) compared to using DIAR (\textasciitilde 0.2\% per step). It demonstrates that our pipeline and the proposed method has great potential in compatible with more accurate speaker probability vectors.

\subsection{Target-Speaker ASR}

\begin{table}[tbp!]
\caption{TS-ASR model evalutations on AMI-IHM eval split}
\vspace{-3ex}
  \label{tab:spk_kernel_tar}
  \centering
  \vspace{2ex}
  \begin{tabular}[c]{rcc|cc|cc}
    \toprule
    \multirow{1}*{\textbf{Speaker}} & \multicolumn{6}{c}{Target-Speaker} \\ 
     \multirow{1}*{\textbf{Encoding}} & \multicolumn{2}{c|}{0 (DIAR)} & \multicolumn{2}{|c|}{0.5} & \multicolumn{2}{|c}{1 (RTTM)}  \\
     \textbf{Scheme} & RTTM & DIAR & RTTM & DIAR & RTTM & DIAR \\
     
   \midrule
    No Spk-Sup.  & 27.70 & - & - & - & - & -  \\
        Projection & 22.64 & \textbf{22.26} & 20.49 & 22.18 & 18.65 & \textbf{25.94} \\
        Sinusoidal & 22.89 & 23.06 & \textbf{19.93} & \textbf{21.74} & 17.43 & 29.57 \\
        Meta-Cat & 22.34 & 22.89 & 20.56 & 22.09 & 18.55 & 27.84 \\ \midrule
        Meta-Cat-R & 23.11 & 23.44 & 20.14 & 22.10 & \textbf{15.01} & 32.92 \\
        Meta-Cat-RP & \textbf{22.24} & 22.78 & 20.63 & 22.83 & 16.31 & 32.94 \\
    \bottomrule
     \end{tabular}
     \vspace{-2ex}
\end{table}

Table~\ref{tab:spk_kernel_tar} compares speaker encoding methods for TS-ASR with different RTTM-Mix-Prob settings. Increasing RTTM-Mix-Prob improves WER under RTTM supervision but degrades it under DIAR, emphasizing the challenge of query-target speaker mapping. Meta-Cat-R achieves the best WER of 15.01\% with RTTM, but it also faces challenges in DIAR setting because the model does not encounter DIAR supervision during training phase. With the help of Meta-Cat-RP, we achieve the best WER of 22.24\% using pure DIAR in training. It shows including all speakers' information can effectively improve the model's robustness towards inaccurate diarization output. 

\begin{table}[tbp!]
\caption{TS-ASR results on LibriSpeechMix. Notation: L - length of auxiliary utterance, SD - speaker deduplication, $^\dag$SA-WER.}
\vspace{-3ex}
  \label{tab:LSmix}
  \centering
  \vspace{2ex}
  \begin{tabular}[r]{rccc|cc}
    \toprule
    \textbf{TS-ASR Systems} & \multirow{1}*{L}  & \multicolumn{2}{c}{\textbf{2-mix }} & \multicolumn{2}{c}{\textbf{3-mix}}  \\ 
    \midrule
        No Spk-Sup. & 7.5 & \multicolumn{2}{c|}{4.30} & \multicolumn{2}{c}{13.26}   \\
        TS-ASR \cite{zhang2023conformer} & 7.5 & \multicolumn{2}{c|}{7.0} & \multicolumn{2}{c}{9.7} \\
        TS-ASR+Spec. \cite{zhang2023conformer} & 7.5 & \multicolumn{2}{c|}{6.3} & \multicolumn{2}{c}{9.0} \\
        E2E SA-ASR \cite{kanda21b_interspeech} & 15 & \multicolumn{2}{c|}{6.8$^\dag$} & \multicolumn{2}{c}{9.6$^\dag$} \\
        E2E SA-ASR SD \cite{kanda21b_interspeech} & 15 & \multicolumn{2}{c|}{6.4$^\dag$} & \multicolumn{2}{c}{8.5$^\dag$} \\
    \midrule
    \textbf{Speaker Encoding} & \multirow{2}*{L}  & \multicolumn{2}{c}{\textbf{2-mix }} & \multicolumn{2}{c}{\textbf{3-mix}}  \\ 
\textbf{Scheme} & & RTTM & DIAR & RTTM & DIAR\\
    \midrule
        Projection  & 7.5 & 8.67 & 9.81 & 28.97 & 29.25 \\
        Sinusoidal  & 7.5 & 5.75 & 5.92 & 17.06 & 17.04 \\
        Meta-Cat    & 7.5 & 4.36 & 5.31 & 9.64 & 10.98 \\
        Meta-Cat-R  & 7.5 & \textbf{3.75} & \textbf{4.36} & \textbf{9.64} & \textbf{10.08} \\
        Meta-Cat-RP & 7.5 & 6.02 & 6.07 & 14.46 & 14.53\\
    \bottomrule
    \vspace{-4ex}
     \end{tabular}
     \vspace{-2ex}
\end{table}
We also evaluate our model on the LibriSpeechMix \cite{kanda2020serialized} dataset using both RTTM and DIAR supervision, as presented in Table~\ref{tab:LSmix}, with the RTTM-Mix-Prob parameter set to 0.5 during training. We follow the same evaluation metric TS-WER used in \cite{zhang2023conformer}. In contrast to real-world datasets like AMI, the results on LibriSpeechMix reveal a minimal disparity between RTTM and DIAR supervision. Our model demonstrates superior performance on the 2-mix subset compared with other state-of-the-art systems. However, for the 3-mix subset, the model exhibits slightly lower performance in comparison to the alternative approaches \cite{zhang2023conformer,kanda21b_interspeech}. 

\subsection{MS/TS-ASR Dual-Task Model}
\begin{table}[tbp!]
\caption{Dual-task model: MS-ASR and TS-ASR evalutations on AMI-IHM test}
  \label{tab:multi-tar}
  \centering
  \vspace{-2ex}
  \begin{tabular}[c]{@{\hspace{1ex}}r@{\hspace{1ex}}|@{\hspace{1ex}}c@{\hspace{1ex}}|@{\hspace{0.85ex}}c@{\hspace{0.85ex}}@{\hspace{0.85ex}}c@{\hspace{0.85ex}}|@{\hspace{0.85ex}}c@{\hspace{0.85ex}}@{\hspace{0.85ex}}c@{\hspace{0.85ex}}|@{\hspace{0.85ex}}c@{\hspace{0.85ex}}@{\hspace{0.85ex}}c@{\hspace{0.85ex}}}
    \toprule
    \multirow{1}*{\textbf{Speaker}}& \multirow{1}*{RTTM}& \multicolumn{4}{c}{Multi-Speaker} & \multicolumn{2}{c}{Target Speaker}\\
    \textbf{Encoding} &\textbf{Mix} & \multicolumn{2}{c}{RTTM} & \multicolumn{2}{c}{DIAR} & RTTM & DIAR \\
    \textbf{Scheme} & \textbf{Prob}& WER & cpWER & WER & cpWER & WER & WER\\
   \midrule 
        No Spk-Sup.& NA & 23.59 & 31.86 & - & -  & 54.98 & -\\
	Projection & 1 & 23.41 & 29.00 & 25.84 & 32.89  & 33.94 & 44.92 \\
	Sinusoidal & 1 & 23.65 & 30.09  & 23.97 & 31.23  & 34.89 & 46.59 \\ \midrule
	\multirow{4}*{Meta-Cat} & 1 & 22.93 & \textbf{26.57}  & 23.68 & \textbf{29.97}  & \textbf{27.05} & \textbf{39.43}  \\
 	 & 0.75 & \textbf{22.77} & 29.09  & 23.09 & 30.47  & 34.63 & 44.12  \\
  	 & 0.5 & 23.05 & 29.68  & 23.25 & 30.82  & 38.39 & 40.95  \\
   	 & 0.25 & 23.15 & 30.69  & \textbf{23.05} & 30.57  & 41.55 & 42.89 \\ 
    \bottomrule
    \end{tabular}
    \vspace{-5ex}
\end{table}

The results for the MS/TS-ASR dual-task model in Table~\ref{tab:multi-tar} show that the Meta-Cat speaker encoding method consistently outperforms other methods in both MS-ASR and TS-ASR tasks. However, the unified, dual-task MS+TS-ASR model shows degraded accuracy compared to models trained exclusively for either MS-ASR or TS-ASR, which is expected due to the additional complexity of handling both speaker differentiation and query-speaker mapping simultaneously. Varying the RTTM-Mix-Prob during Meta-Cat training reveals that increasing the use of RTTM supervision improves performance in the MS-ASR task, but the trend is less clear for the TS-ASR task, where higher RTTM-Mix-Prob values do not consistently improve accuracy when tested with DIAR supervision. This discrepancy likely stems from the difference in training objectives between the dual-task model and a dedicated TS-ASR model, indicating that the dual-task model may struggle more with extracting the target speaker’s transcription from diarization predictions.

\section{Conclusions}
\label{sec:conclusion}
In this study, we proposed novel frameworks to support multi-speaker (MS) and target-speaker (TS) functions within an automatic speech recognition (ASR) model based on a single architecture. We introduced Meta-Cat and its variants, Meta-Cat-(R, RP), a new speaker information encoding scheme that effectively incorporates speaker supervision from an integrated speaker diarization model. Meta-Cat demonstrated superior performance compared to other speaker information encoding methods on MS-ASR and TS-ASR tasks. Additionally, although it showed degraded performance, a unified dual-task model was presented, indicating the potential for multi-task capabilities within a single model. Further research is needed on training schemes to enable robust dual-task capability, such as employing adapters or multi-head architectures.

\vfill\pagebreak



\bibliographystyle{IEEEtran}
\bibliography{strings,refs,reference}
\end{document}